\documentclass[aps,prl,amsmath,amssymb,reprint,superscriptaddress,longbibliography]{revtex4-1}
\usepackage{graphicx}
\usepackage{amsmath}
\usepackage{dcolumn}
\usepackage{bm}
\usepackage{bbold}
\usepackage{color}
\usepackage{amsfonts}
\usepackage{amssymb}
\usepackage{mathrsfs}
\usepackage{tabularx}
\usepackage{braket}
\usepackage{mathtools}
\usepackage{soul}			

\usepackage{caption}
\usepackage{subcaption}

\begin{document}

\title{Reconfigurable Microwave Photonic Topological Insulator}

\author{Maxim Goryachev}\affiliation{ARC Centre of Excellence for Engineered Quantum Systems, School of Physics, University of Western Australia, 35 Stirling Highway, Crawley WA 6009, Australia}
\author{Michael E. Tobar}\affiliation{ARC Centre of Excellence for Engineered Quantum Systems, School of Physics, University of Western Australia, 35 Stirling Highway, Crawley WA 6009, Australia}

\date{\today}

\begin{abstract}

{Using full 3D finite element simulation and underlining Hamiltonian models,} we demonstrate reconfigurable photonic analogues of topological insulators on a regular lattice of tunable posts in a re-entrant 3D lumped element type system. The tunability allows dynamical {\it in-situ} change of media chirality and other properties via {\color{black}alteration of the same parameter for all posts}, and as a result, great flexibility in choice of bulk/edge configurations. Additionally, one way photon transport without an external magnetic field is demonstrated. The ideas are illustrated by using both full finite element simulation as well as simplified harmonic oscillator models. {\color{black}Dynamical reconfigurability} of the proposed systems paves the way to a new class of systems that can be employed for random access, topological signal processing and sensing.


\end{abstract}

\maketitle


\section{Introduction}

Ideas on nontrivial topological properties of various systems can be found in many areas of research such as nonlinear dynamics\cite{topol1}, control system theory\cite{Vakhrameev:1995aa}, biology\cite{topol2}, data analysis\cite{td1}, etc. In physics and material science, these ideas have found applications to the so-called topological insulators, which have recently drawn considerable attention due to their unusual properties. As this type of condensed matter system is not very abundant in nature, many researchers invoke metamaterials and metastructures as a powerful method of engineering new materials with required properties\cite{veselago}. Moreover, metamaterials are typically photon systems that help to overcome many limitations of the condensed state of matter. Some examples of such photon meta systems with nontrivial topological phases are hyperbolic metamaterials in the long wave length limit\cite{Gao:2015aa}, metawaveguides and metamaterials\cite{Khanikaev:2013aa,Ma:2015aa,Slobozhanyuk:2016aa,Cheng:2016aa}, linear photonic systems and resonant structures\cite{HafeziM.:2013aa,Pasek:2014aa,Slobozhanyuk:2015aa}, coupled optical\cite{Umucallar:2011aa,Liang:2013aa} and microwave\cite{Anderson2016,Petrescu:2012aa} cavity arrays or networks\cite{Hu:2015aa} as well as acoustical systems\cite{Xiao:2015aa, Yang:2015aa} and circuits\cite{Ningyuan:2015aa}. Properties of such materials regardless of their particular realisations have been also studied\cite{Zuo:2013aa,Lumer:2013aa,Fulga:2012aa,Hafezi:2014aa}.

In this work, we propose a new concept of realisation of microwave photon analogues of topological insulators using tunable simple and non chiral objects defined on regular grids. The novelty of this approach is based on the proposal to utilize dynamically tunable re-entrant multipost structures and special arrangement of these posts that results in nontrivial topological properties. Moreover, the high tunability of these objects makes it possible to reconfigure the system topology {\it in situ} by changing the global parameters of media such as handedness of constituent objects, position of boundaries between parts with different chirality, etc. A reconfigurable topological metastructure has been recently demonstrated\cite{Cheng:2016aa}, although its realisation requires external intervention into internal structure properties. Unlike this previously proposed metastructure\cite{Cheng:2016aa} that can not be reconfigurable dynamically, in the approach presented in this work, reconfigurability can be achieved with piezoactuators or even electronically with DC voltage variable capacitors. This valuable property is achieved in this work by implementing the re-entrant 3D lumped mode approach. 

{For a re-entrant cavity, the resonant mode obtains both inductance and capacitance from a post (or metalic rod) attached at one end to a closed cavity and with a gap between the post and the cavity surface at the other end\cite{reen2}. The inductance is determined by the resonant magnetic field circulating the post, while the high tunability is made possible due to the very dense concentration of the electric field in the small gap under the post, forming a capacitor\cite{reen1}. This capacitor may be made highly tunable by varying the electrical path length within the gap.} 
This re-entrant mode approach was recently expanded to a multi post system\cite{Goryachev:2015aa,Goryachev:2015ab} to enhance coupling to spin systems\cite{Goryachev:2014aa,Creedon:2015aa} and engineer the microwave response of complex spin-cavity systems\cite{Kostylev:2016aa}. Besides these proposals, the re-entrant types of cavities have been used for many applications including chemistry\cite{enhchem}, material science\cite{Harrington}, plasma\cite{Hemawan:2009aa} and accelerator\cite{Ceperley:1975aa} physics, as well as parametric transducers\cite{tobar1993parametric}. 

It should be emphasized that the system exploited in this work is essentially different to realizations of photonic topological insulators considered before\cite{Khanikaev:2013aa,Ma:2015aa} as they exploit different types of modes. In the present work, all modes are based on re-entrant types of resonances\cite{reen2} (essentially lumped modes in the sub-wavelength limit) in contrast to the typical Transverse Electric or Transverse Magnetic modes in a cavity resonator\cite{Khanikaev:2013aa,Ma:2015aa}. Moreover, in previous work, interfaces are made between topological media constructed only based on elements of explicitly different position or dimensions that can not be tuned onto into another. In the present case, the left and right type media are all identical square lattices of the same objects that can be tuned into one another via a simple gap change of all or several posts.  

The system is functionality demonstrated using a Hamiltonian model of the re-entrant post lattice and verified by full scale Finite Element Model (FEM) simulations with COMSOL, with the results explained in terms of cavity design. All effects presented in the following sections are observed for different dimensions of the simulated structures.

\section{Chiral Post Molecules}

The concept of tuneable microwave media based on arrangement of re-entrant posts has been introduced and discussed for both 1D\cite{Goryachev:2015aa} and 2D\cite{Goryachev:2015ab} lattices. {These lattices are formed by translation of metallic re-entrant posts in a closed conducting cavity. Each, typically circular, post is characterised by a gap between its end and the opposite cavity wall. This gap forms a capacitor that together with post inductance form a re-entrant mode\cite{reen0,reen1}. Such modes are characterised by the electric field pointing along the post direction (referred to as $z$ axis thought this work) and compactly located in the post gap and the magnetic field lying in the cavity plane normal to posts ($x-y$ plane). Due to the compact location of the electric field, the resonance frequency is found to be highly tunable\cite{reen1} via changes in the gap height. It has been demonstrated\cite{Goryachev:2015aa} that cavities with multiple posts exhibit multiple re-entrant resonances (number of resonances correspond to the number of gaps) with wave distributions defined on post lattices. For regular lattices, these modes can be classified as optical and acoustic exhibiting typical band structures with band gaps.}

{An important feature of the re-entrant cavity approach is the fact that cavities of the same resonant frequency may be implemented with completely different geometrical dimensions. Indeed, a resonance frequency of a stand alone post depends primarily on the gap capacitance, that is a function of the post diameter and the gap height, and the post inductance, that primarily depends on the post length and its diameter.
Thus, there are three free parameters that can be tuned to achieve the same resonance frequencies. Moreover, the most important parameter is the gap height that might be used to tune the cavity over a very large frequency range. In this respect, the re-entrant cavity has much in common with lumped component resonant tanks where the same resonance frequencies can be achieved with different geometric dimensions of the underlying lumped components.}

In the case of 2D lattices\cite{Goryachev:2015ab}, the concept of 'microwave post molecules' was demonstrated. In this context, by arranging re-entrant posts in close subgroups, one can create lattices of such groupings with lattice constants larger than a typical distance within the molecule. Within the lattice, the closer arrangement of posts within the molecule makes the internal inter-post coupling larger than that between posts in different molecules. As a result, the energy required to change the mode structure within a molecule, given by current directions at an instance of time, is much lager than that required for energy change on the larger lattice scale between the molecules themselves. This scale difference gives rise to different mode families and energy band gaps\cite{Goryachev:2015ab}. { Given a lattice of closely located groups of molecules, the structure exhibits a number of types of modes corresponding to the number of posts within a molecule. Each mode family is characterised by the combination of  directions of the electrical field $E_z$ under each post in a given instance of time. The lowest energy mode is always represented by electrical field vectors pointing in the same direction, i.e. sequences of the form $\{\uparrow\uparrow..\uparrow\}$, whereas the highest energy mode is always associated with the largest number of electrical field alternations within a molecule giving $\{\uparrow\downarrow..\downarrow\uparrow\}$ sequences. Particular field direction distributions depend on molecule symmetries and might be thought of as wave polarisations\cite{Goryachev:2015ab}. Each mode family, or equivalently a type of polarisation, forms a band of harmonic waves with a certain propagation speed.}

The same idea of post molecules may be used to implement chirality. Indeed, it is possible to arrange $N$ posts with different dimensions into molecules with clear handedness. One possible way of doing that is to make a circular array of posts with one or more changing post parameters. In practice, the easiest parameter to vary for re-entrant systems is the post gap, i.e. the distance $h$ between the post and the opposite wall. In principle, this parameter can be tuned mechanically or electronically (via a varactor, for example) resulting in change of equivalent capacitance. 

\begin{figure}[htbp]
\centering
\includegraphics[width=1\columnwidth]{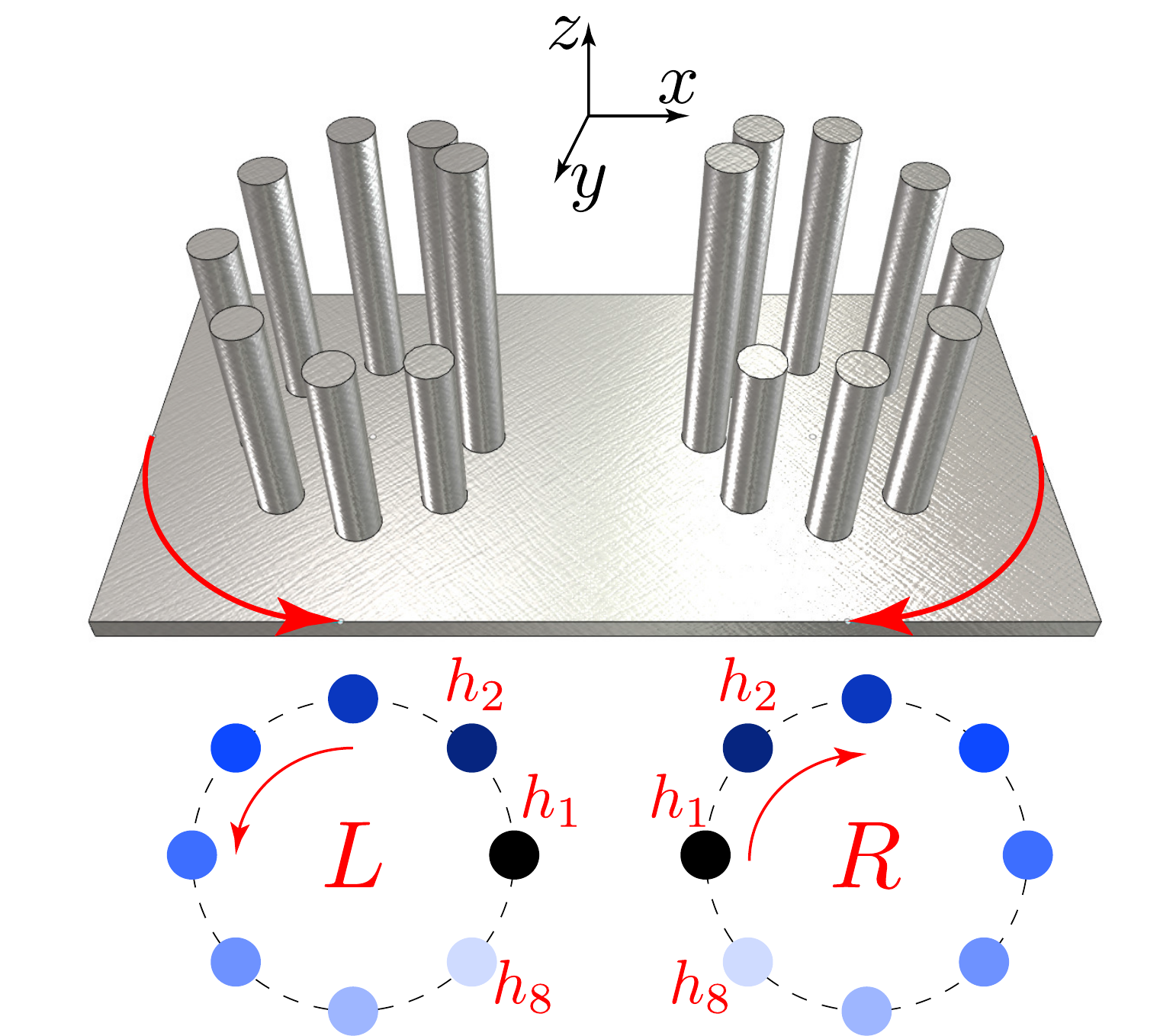}
\caption{Left and right handed circular $N=8$ post molecules with circular gap variation. Gap changes $\Delta h$ are exaggerated to facilitate visibility. Side and upper cavity walls are not shown.}
\label{LR}
\end{figure}

An example of the molecules with handedness is shown in Fig.~\ref{LR}, where each molecule has $N = 8$ posts, and the arrangement may be referred to as either right or left handed. Here each post has individual gap $h_i$ { between the post end and the upper cavity wall}, which increases with the post position $i$ changing clockwise or counter-clockwise. The structure shown in Fig.~\ref{LR} is completed to a closed multipost cavity by introducing side and upper conducting walls. As well the structure can be translated in the $x-y$ plane to make a lattice of molecules and only then completed by the remaining side walls. The $N=8$ post molecules shown in Fig.~\ref{LR} may exhibit eight types of polarisations spanning from $\{\uparrow\uparrow\uparrow\uparrow\uparrow\uparrow\uparrow\uparrow\}$ giving the lowest energy band to $\{\uparrow\downarrow\uparrow\downarrow\uparrow\downarrow\uparrow\downarrow\}$ for the highest energy band.

It should be mentioned that terms 'right' and 'left' have no particular meaning for the types of molecules considered in this work. They are employed simply to distinguish the two types of handedness. Terms 'clockwise' and 'anticlockwise' could be used on the same grounds. Although, it is important to set the direction of post change to avoid confusion. In this work, we set the direction from $h_i$ to $h_j$ where $i>j$. 

There is an alternative way of implementing chirality into re-entrant post media that is to make posts chiral themselves. This could be done by designing spiral like posts with two possible ways of winding. Although, in practice this is significantly more difficult to implement. Additionally, this approach would not allow {\it in-situ} reconfiguration of the molecules and thus the media.

\section{$N=4$ Circular Molecules}
\label{this}

For all further numerical examples, we use the case of $N=4$ circular post molecules (i.e. square shaped arrangement). This choice is made for symmetry reasons, so that it is easy to arrange into square symmetric lattices. Moreover, for the same reasons we consider a limiting case in which the lattice constant equals twice the dimension of the molecule square, so that the distances between posts within a molecule and with its next neighbours are equal.

As it is discussed in the previous section, posts within a molecule are different only by their gaps. This fact makes the medium bulk anisotropic. In one direction, the medium properties are set by alternating chains $h_4-h_1$ and $h_3-h_2$ and in other by $h_4-h_3$ and $h_1-h_2$. 

{ For example, let us consider an isolated $N = 4$ molecule with angular frequencies associated with each posts distributed clockwise as $(\omega_0- 2\Delta,\omega_0 - \Delta,\omega_0+\Delta,\omega_0+2\Delta)$ and only nearest neighbour interactions.
The Hamiltonian of this molecule can be written in a rotating frame as:
\begin{equation}
\begin{array}{l}
\displaystyle H = (2\Delta a^\dagger_{1} a_{1} + \Delta a^\dagger_{2} a_{2} -\Delta a^\dagger_{3} a_{3} - 2\Delta a^\dagger_{4} a_{4}  \\
\displaystyle +g_n (a^\dagger_{1}a_{2}+a^\dagger_{2}a_{3}+a^\dagger_{3}a_{4}+a^\dagger_{4}a_{1})+\text{h.c.}
\label{Ham0}
\end{array}
\end{equation}
where $a^\dagger_{i}$ ($a_{i}$) are creation (annihilation) operators for Harmonic oscillators representing each post and $g_n$ is coupling between nearest neighbour posts.\ 
In this case, the anti-diagonal matrix $\text{antidiag}(C) = (-1,1,-1,1)$ has the following properties:
\begin{equation}
\begin{array}{l}
\displaystyle H = - C H^T C^{-1}, C^\dagger C = 1, C^T = -C.
\label{proper}
\end{array}
\end{equation}
These properties allows us to interpret this matrix as an operator that anti-commutes with the Hamiltonian or as a realisation of particle-hole symmetry \cite{Schnyder:2008aa}. By swapping the order of posts from clockwise to anticlockwise, the L molecule Hamiltonian is turned into a R molecule Hamiltonian, implying that one type of molecule is the antiparticle of the other. 
Additionally, the molecule is time reversal symmetric that together with the particle-hole symmetry suggests interpretation of the system as chiral. 
}


It is important to note that in such limiting case, there is a mirror symmetry between the molecules interior and some portions of the intermolecular space. By exchanging these parts, one switches from one type of handedness to another. But as long as this transformation is done for both types of media, the duality between left and right types of molecules is preserved. 

\subsection{Media Interfaces}

\begin{figure}[htbp]
\centering
\includegraphics[width=1\columnwidth]{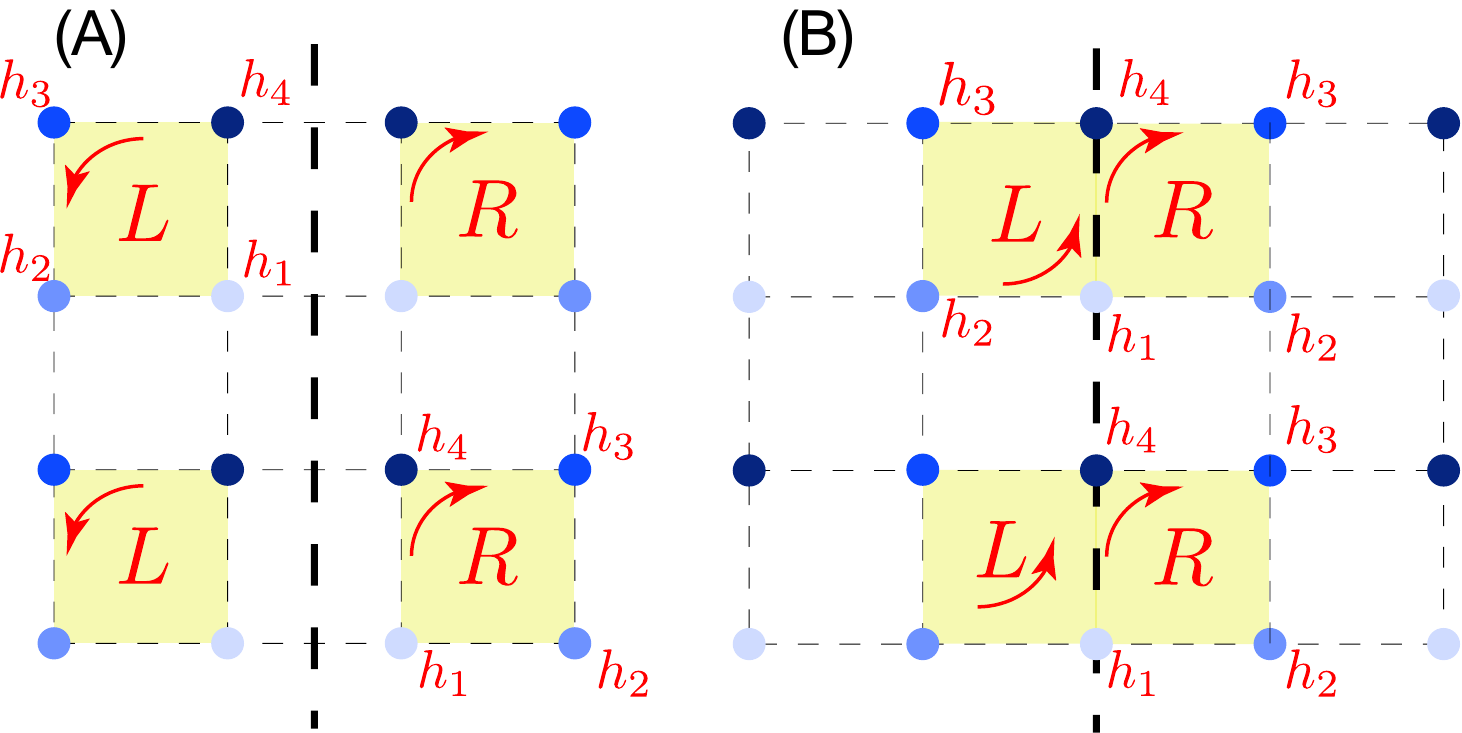}
\caption{Two types of interfaces between left and right handed latices: (A) interspace, (B) edge. Shaded areas denote interiors of four post molecules. }
\label{inters}
\end{figure}

In this section we investigate the surface modes that can be found on the interface between lattices of two different chiralities. Assuming a regular square grid of posts, there are two possible interfaces: interspace and edge interfaces as shown in Fig.~\ref{inters}. The former is a situation when the interface is within the inter-molecule space, while the latter is a result of the interface coincident with one lattice edge. These two cases are shown in Fig.~\ref{inters}(A) and Fig.~\ref{inters}(B) respectively. Given an ordered set of numbers $\{h_4,h_3,h_2,h_1\}$ in a molecular cell, 16 combinations preserving the lattice structure are possible, created by four possible rotations on each side. Four of these interfaces preserve mirror symmetry around the interspace interface (as shown in Fig.~\ref{inters}(A)) and are considered further in our modelling. In contrast, the edge interface shown in Fig.~\ref{inters} (B) has 4 combinations, in which all of those also preserve the mirror symmetry.

The other type of interface is formed between the lattice and side walls of the cavity. It is known in electro-magnetics that ideal conductor may be considered as an ideal mirror. Thus, the overall system may be considered as the real lattice and its virtual reflection resulting in periodic boundary conditions. So, the situation is reduced to the mirror symmetry preserving interspace interface considered shown in Fig.~\ref{inters} (A) and should give rise to edge modes as well.

\begin{figure*}[htbp]
\centering
\includegraphics[width=2\columnwidth]{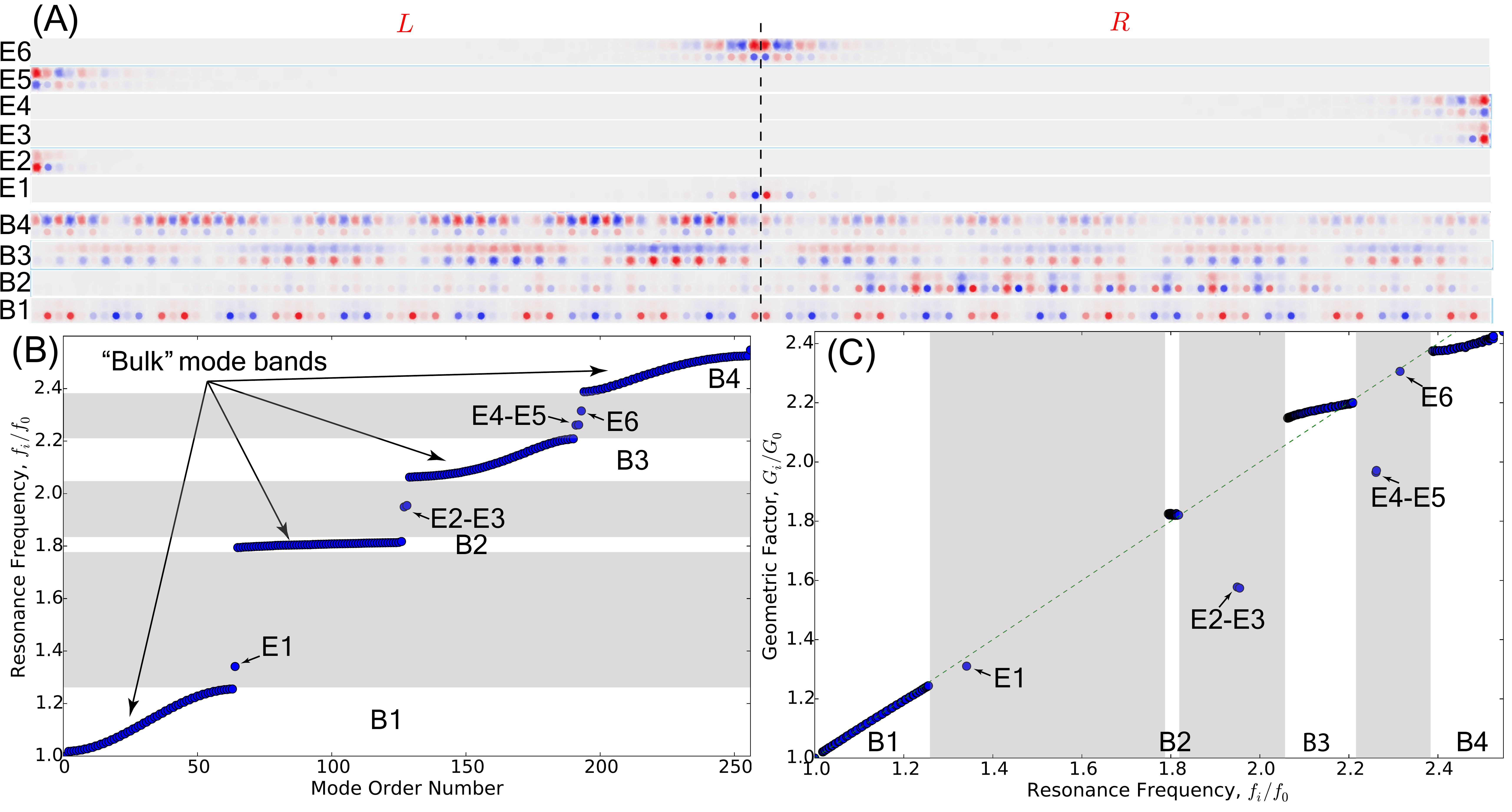}
\caption{(A) Color plot of the distribution of the electric field {($E_z$ component)} directly under the gaps for the 32 left handed ($L$) and right handed ($R$) square molecules, depicted for edge modes E1 to E6  and bulk modes from bands B1-B4, the vertical dashed line illustrated the boundary between the $R$ and $L$ handed molecules. (B) Distribution of the mode frequencies in the chiral structure, where the shaded areas show band gaps for 'bulk' modes. (C) Scaled Geometric Factor ratio for all system modes as a function of scaled resonance frequency.
}
\label{disper1}
\end{figure*}

\section{One Dimensional Implementation}

To simplify numerical simulation requirements, we begin with a quasi one dimensional system. In this model, a single chain of square molecules is implemented. The chain is divided into left and right handed halves. From the observer point of view the system consists of two strings of posts separated by the same distance as posts within the same string. The only difference between these strings is gaps under the posts.

For the numerical example we take a chain of 32 left-handed and 32 right handed $N=4$ square molecules. The result of the numerical simulation is shown in Fig.~\ref{disper1}, which presents electric field density mode plots of typical mode structures (A), distribution of modes and bands over frequencies relative to the first eigenfrequency (B) and dependence of the geometric factor on the mode frequency (C).

Fig.~\ref{disper1} demonstrates existence of four 'bulk' mode bounds. Each bound is associated with modes of certain molecule polarisation characterised by a certain pattern of current directions within a molecule\cite{Goryachev:2015ab}: B1, all four currents point in the same direction; B2, one post current is opposite to three others; B3, one post current is opposite to three others; B4, two post currents are opposite to two others. Modes of types B2 and B3 are not degenerate due to high anisotropy of chiral molecules considered in this work. Between these groups of mode, the system exhibits three band gaps. The band gaps are not completely empty of modes, they exhibit a limited number of eigenfrequencies for which the mode distribution can not be classified belonging to any of described bands. The gap between B1 and B2 contains one mode E1 which shows energy concentration only on the interface (few bordering molecules) between $L$ and $R$ parts of the chain (Fig.~\ref{disper1} (A), E1). The distribution of the electric field across the interface is antisymmetric. The symmetrical distribution is observed for the mode E6 existing between B3-B4.
The gap B2-B3 contains two modes for which the energy is concentrated on one of the interfaces between the chain and the side walls (Fig.~\ref{disper1} (A), E2 and E3). The frequencies are not completely degenerate due the asymmetry of the two types of molecules. Similar picture is seen for modes E4 and E5 existing between B3 and B4. 

{Besides the resonance frequency, another important parameter for the cavity design and optimisation is the Geometric factor, $G$ in ohms. This parameter characterises cavity losses due to conducting walls without any assumptions on their actual conductance or surface resistance $R_s$, and thus makes it equivalent to the Quality factor up to a material constant, where $Q = G/R_s$. Thus, it is important to compare this parameter between bulk and edge modes.} The Geometric factor for a mode with angular frequency $\omega_i$ measured in units of resistance is given by the relation:
\begin{equation}
\begin{array}{l}
\displaystyle G_i = \omega_i \int_V |H|^2 dv/ \int_S |H|^2 ds,
\label{GF}
\end{array}
\end{equation}
where two integrals of squares of magnetic field intensity $H$ are calculated over the total system volume $V$ and conducting surfaces $S$. { In other words, this factor relates the portion of energy stored in the bulk (free space) to the energy dissipated in the currents at the conducting surfaces. For calculation purposes, it is sufficient to assume the conductors are perfect to determine the tangential components of $H$, as for any resonant system, this will be a good approximation.} From Fig.~\ref{disper1} it is seen that $G_i/G_0$ tend to group near the $G_i/G_0 = f_i/f_0$ line {due to the $\omega_i$ factor in Eq. (\ref{GF}). Edge modes on the interface between the two media (E1 and E6) are also found in the vicinity of this line.}
 The only four significant exceptions are represented by the modes on the outer edges (edge modes E2-E5). These modes couple more strongly to the cavity walls giving higher surface component of the energy, and hence will have lower geometric factor and consequently $Q$-factor.

\begin{figure}[!th]
\centering
\includegraphics[width=1\columnwidth]{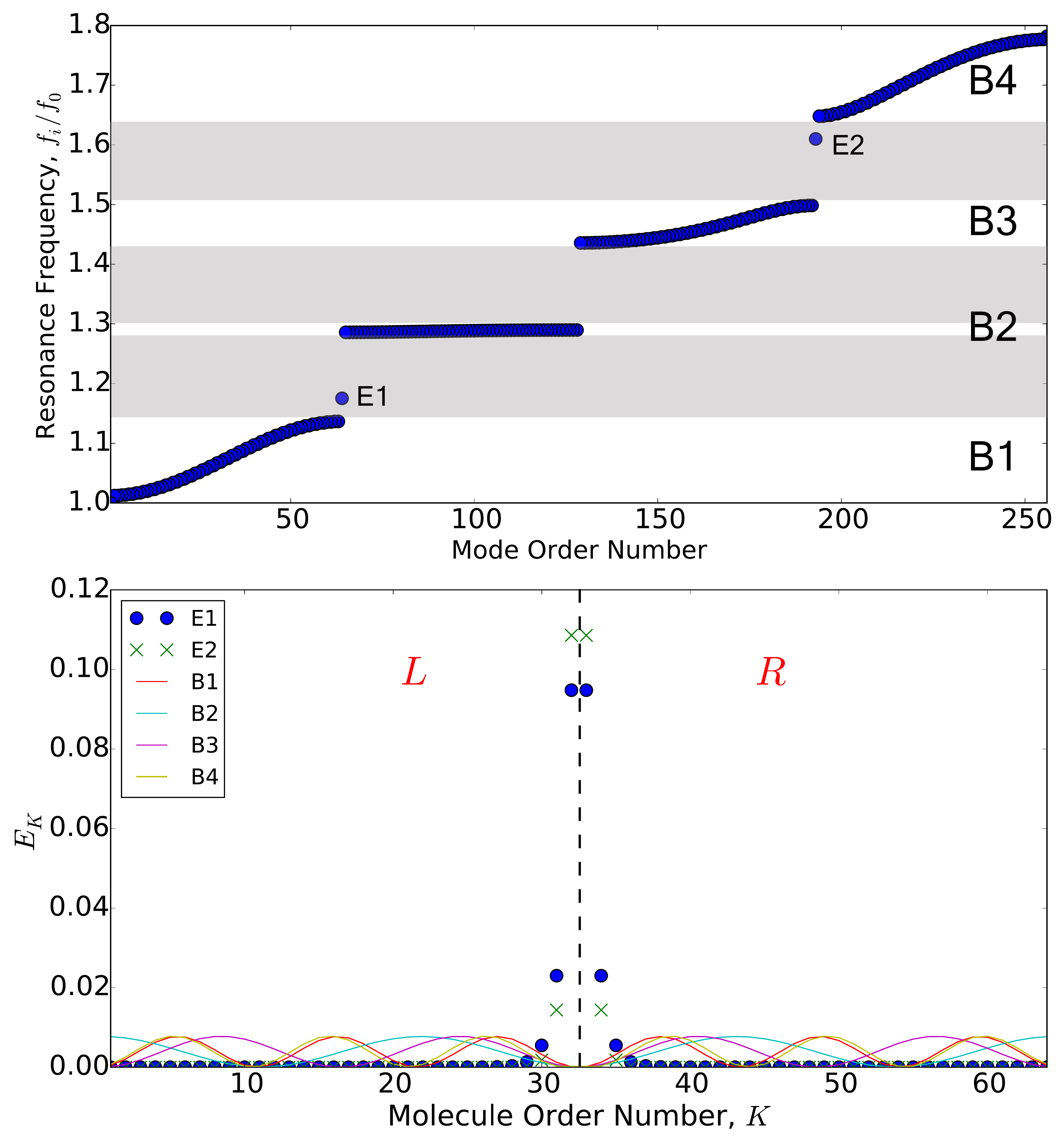}
\caption{(A) Distribution of the eigenfrequencies for the chain square lattice consisting of two halves of different chirality.  (B) Dependence of the energy stored in each molecule on its number $K$ in the chain for edge and examples of 'bulk' modes from each band.
}
\label{top1}
\end{figure}

The same result can be obtained with a Harmonic Oscillator (HO) model. In this model each post is associated with a HO whose angular frequencies $\omega_{i,K}$ are distributed according to the logic explained above. For simplicity 
The system Hamiltonian may be written as follows:
\begin{equation}
\begin{array}{l}
\displaystyle H = \sum_{K=1}^M\Big(\sum_{i=1}^N\omega_{i,K} a^\dagger_{i,K} a_{i,K} +g_n\sum_{i=1}^{N-1}a^\dagger_{i,K}a_{i+1,K}+\\
\displaystyle+g_c\sum_{i=1}^{N-2}a^\dagger_{i,K}a_{i+2,K}+\text{h.c.}\Big) + \sum_{K=1}^{M-1}\Big(g_n(a^\dagger_{2,K}a_{1,K+1}+\\
\displaystyle a^\dagger_{4,K}a_{3,K+1})+g_c(a^\dagger_{2,K}a_{3,K+1}+a^\dagger_{4,K}a_{1,K+1})+\text{h.c}\Big)
\label{Ham1}
\end{array}
\end{equation}
where indices $i$ and $K$ refer to the post index inside a molecule and a molecule index in the chain, $g_n$ and $g_c$ are couplings between nearest neighbour  posts and posts across a unit cell respectively, $a^\dagger_{i,K}$ ($a_{i,K}$) is a creation (annihilation operator) for post $(i,K)$. Also, $\omega_{i,K}$ dependence on $i$ has two distributions: $\omega_1>\omega_2>\omega_3>\omega_4$ corresponding to left chiral molecules ($K<N/2$) and $\omega_1<\omega_2<\omega_3<\omega_4$ corresponding to right chiral molecules ($K>N/2$). The result of eigenvalue decomposition of this Hamiltonian is shown in Fig.~\ref{top1} where the numerical parameters are chosen to match the full FEM results qualitatively.

Figure~\ref{top1} (A) confirms existence of the four bands B1-B4 as well as edge states in corresponding band gaps. The high concentration of the energy near the $R/L$ media interface is seen from plot (B) where nonzero components of corresponding system eigenvectors concentrate near the interface compared to examples of the bulk modes from each band. These components are averaged over for each elementary cell of the chain meaning that unlike the FEM simulations every data point represents a molecule rather than a post. The energy distribution is calculated as a sum of squared elements of eigenvectors distributed on each molecule giving average molecule excitation:
\begin{equation}
\begin{array}{l}
\displaystyle E_K = \sum_{i=1}^Nv_{i,K}^2,
\label{energy}
\end{array}
\end{equation}
where $v_{i,K}$ is a value of the eigenvector on the $i$th post of the $K$th molecule.

The main qualitative difference between FEM and Hamiltonian models are absence of edge modes at the ends of the chain for the latter. This discrepancy is explained by the fact that the FEM model effectively imposes periodic boundary conditions while Hamiltonian model is written with fixed boundary condition. The fixed conditions cannot be viewed as another interface between two types of media giving no edge modes.

\section{Two Dimensional Example}

\begin{figure*}[!b]
\centering
\includegraphics[width=2\columnwidth]{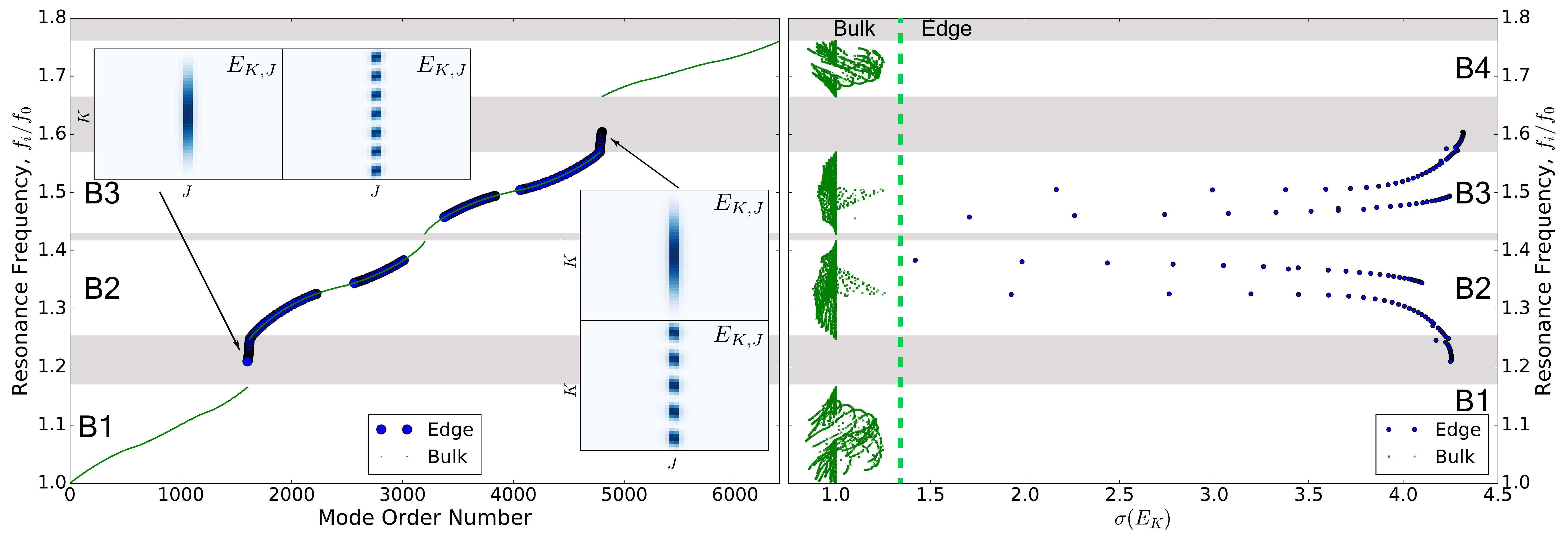}
\caption{(A) Distribution of the eigenfrequencies for the square 2D lattice  constructed of left and right handed molecules. (B) Deviation of the mode energy from the value at the media interface. The insets show energy distribution $E_{K,J}$ for several examples of the edge modes.
}
\label{top2}
\end{figure*}

For the two dimensional example, we implement the same square molecule lattice as described in previous sections. In the 2D case molecules extend in both directions forming a regular grid. This lattice is divided into halves representing left and right chiral media. The border lies along the spacial coordinate $J$ and normal to the second coordinate $K$. As before, both coordinates represent the number of a post molecule in the lattice. An extended 2D version of Hamiltonian~(\ref{Ham1}) can be written in this case whose eigenmodes are identified in the similar fashion. The corresponding results are shown in Fig.~\ref{top2}.  
Due to existence of the spacial extension of the edge between media (the edge is an 1D system on its own), the edge modes form their own bands rather than being single modes in the band gaps of bulk modes as the case of a chain. So, in order to separate two kinds of modes, one can calculate the deviation of the mode energy in the $K$ direction:
{
\begin{equation}
\begin{array}{l}
\displaystyle \sigma^2(E_{K,J}) = \sum_{K,J=1}^{N_K,N_J}(E_{K,J}^2-E_{K,N_J/2}^2)/ \sum_{K,J=1}^{N_K,N_J}E^2_{K,J},
\label{devi}
\end{array}
\end{equation}
where $N_J$ and $N_K$ are the total number of molecules along the corresponding directions, $E_{K,N_J/2}$ characterize the energy at the media interface. }
Bulk modes are characterised by a small energy {deviation due to almost harmonic wave solutions in the balk and on the edge}, while edge modes with their sharp peak result in considerably large {deviation as values of $E_{K,J}$ decay to zero in the bulk}. This idea can be seen in Fig.~\ref{top1} (A) where the sharp distribution of the edge mode energy is apparent. The calculated dispersion for each mode from Fig.~\ref{top2} (B) along $K$ is shown in (B) where all data points are classified in two groups according to the value of the dispersion $\sigma$. This gives two kinds of mode bands in (A) representing bulk and edge modes. The insets show examples of the edge mode energy distributions { $E_{K,J}$ that is a generalisation of Eq.~\ref{energy} on the 2D lattice of $N=4$ molecules.}


\begin{figure}[htbp]
\centering
\includegraphics[width=1\columnwidth]{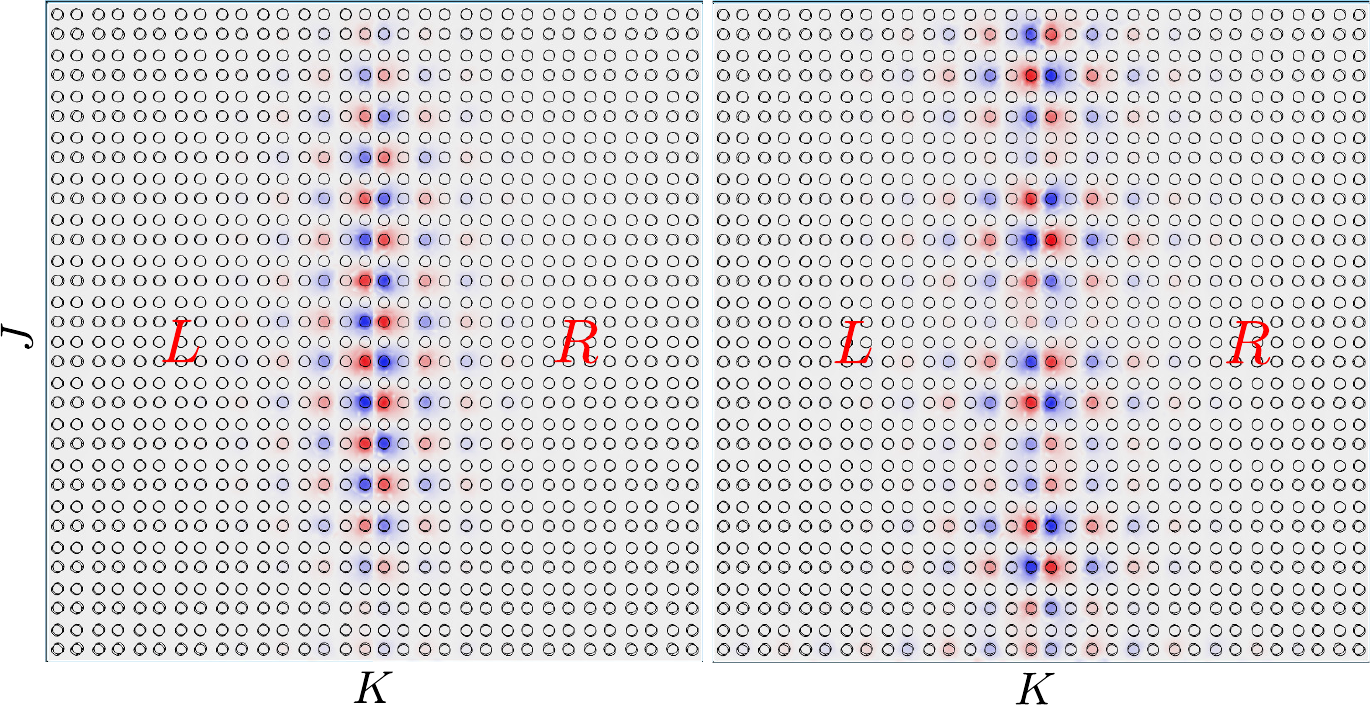}
\caption{Electric field distribution {($E_z$ component)} for first and fourth order edge modes arising on the surface between $R$ and $L$ materials. Every circle represents a post.
}
\label{2Dfull}
\end{figure}

The results of the Hamiltonian model presented in Fig.~\ref{top2} are confirmed with the full FEM modelling of the post system. The result shown in Fig.~\ref{2Dfull} is presented in terms of the electrical field along the posts in their gap plane. Note, that unlike in the HO model, here, every circle represents a post but not a molecule. The result confirms existence of a whole band of edge modes existing on the boundary whose first and second harmonics are shown in this figure. Also, the results demonstrate existence of the modes on the border between the lattice and the cavity wall similar to the 1D case as it is expected from the periodic boundary conditions.

\begin{figure}[htbp]
\centering
\includegraphics[width=1\columnwidth]{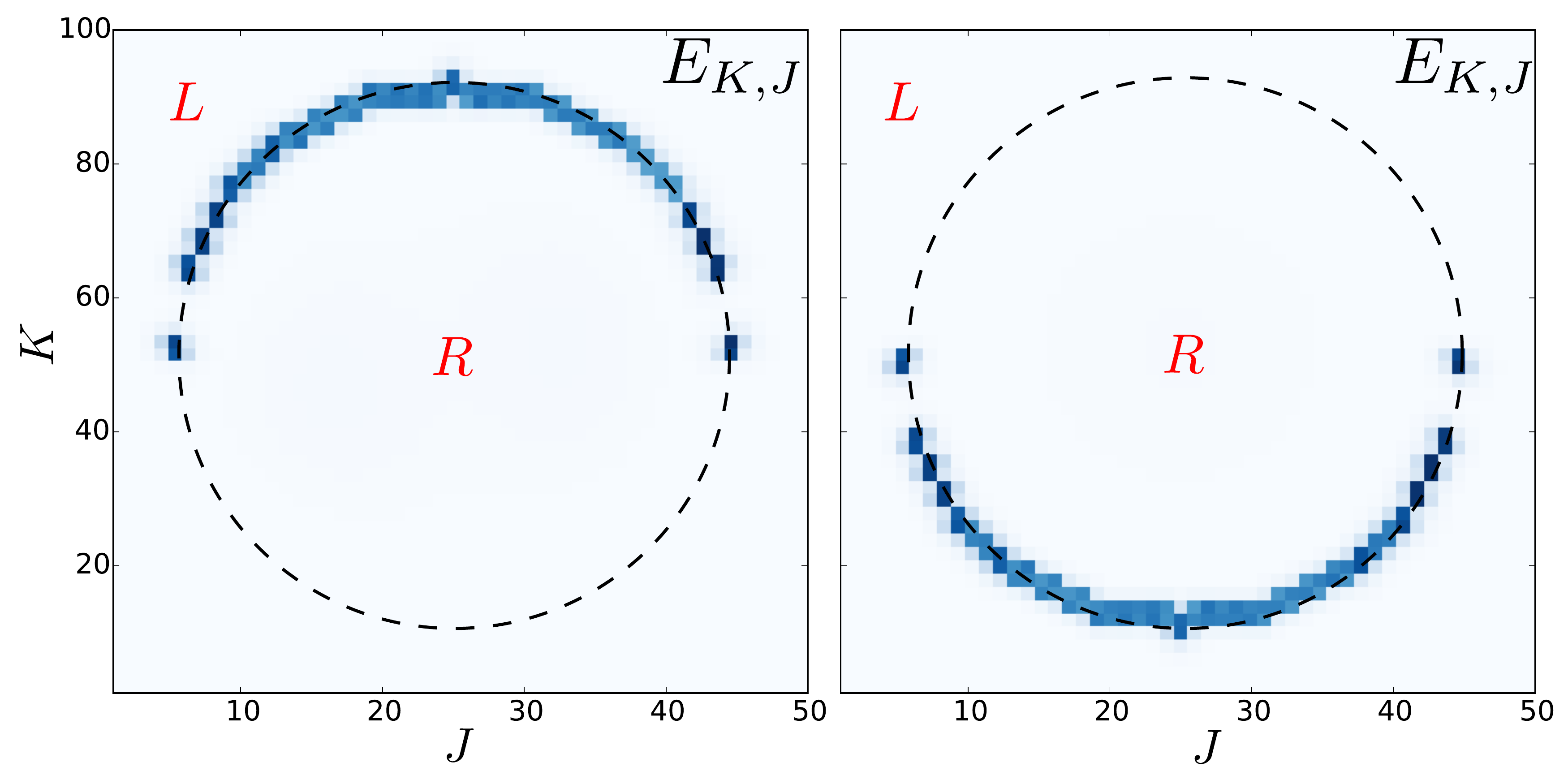}
\caption{Energy distribution $E_{K,J}$ for low order edge modes arising on the interface between a disk of material $R$ enclosed into material $L$ both defined on a regular grid. The dashed curve shows the border between two media.
}
\label{topC}
\end{figure}

Due to high tunability of each particular gap in the array, one may form other geometries on the same 2D regular square grid. For instance, it is possible to define circle area of right handed material enclosed in left handed environment. Energy distribution for some edge modes arising on the border between two materials is shown in Fig.~\ref{topC} that depicts a few edge modes on the circle boundary. This results demonstrates the flexibility of the proposed method: one can define and adjust any topology on the regular 2D grid. 

\section{One Way Photon Transport}

\begin{figure}[ht!]
\centering
\includegraphics[width=1\columnwidth]{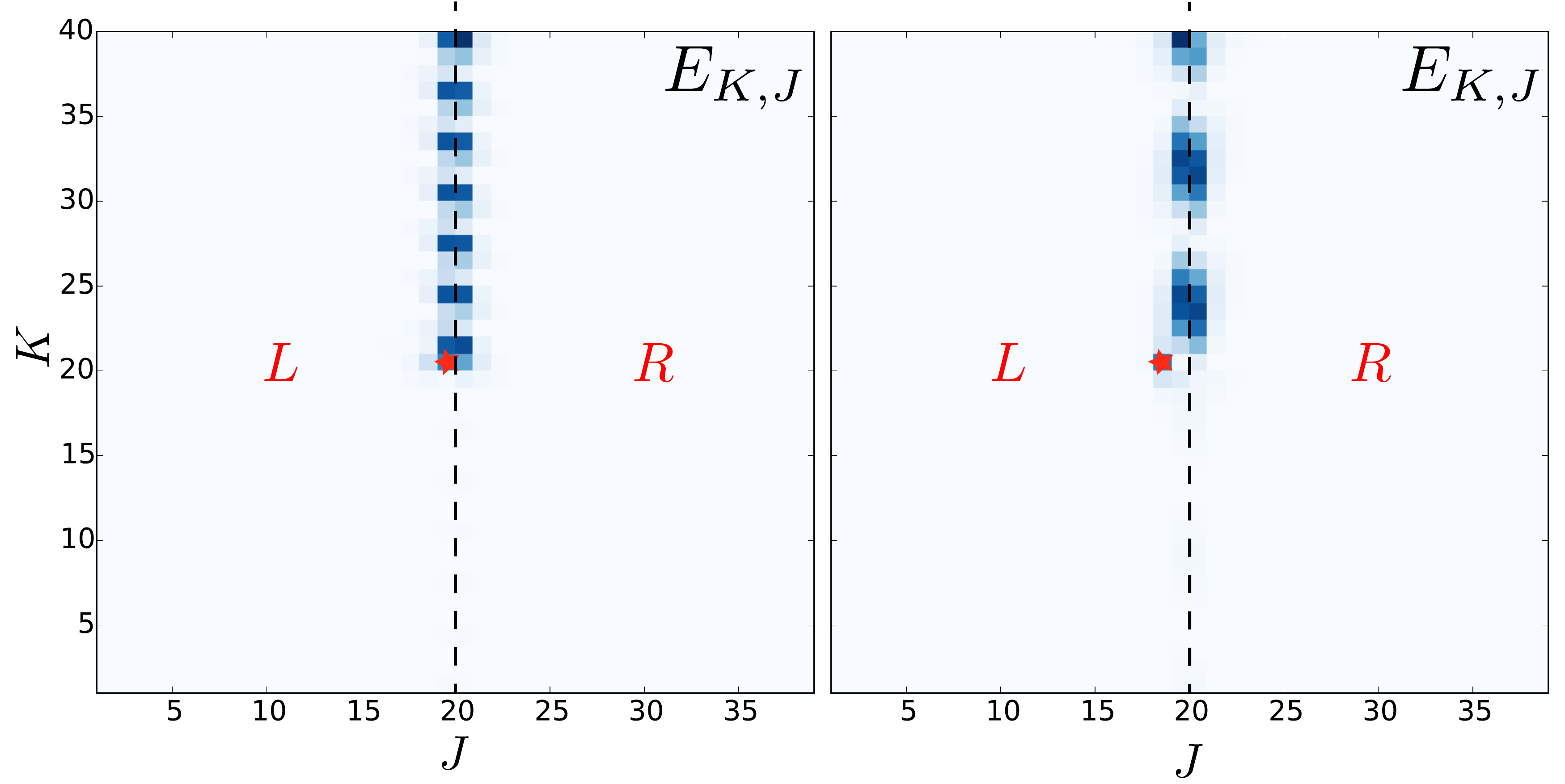}
\caption{Energy distribution $E_{K,J}$ for edge modes demonstrating appearance of the one way photon transport in a lattice with an external signal input. The star signifies the entry port molecule.
}
\label{topTS}
\end{figure}

{\color{black}Another intriguing feature of photonic topological insulators is the ability to design a system with one-way transport. This is difficult to achieve in a condensed matter system because of the bulk-boundary correspondence, due to the fact that a finite system hosts counter-propagating one-way states on each edge of a sample and therefore there is no net transport. This has only recently been circumvented in the composite metal on topological insulator system\cite{Foa-Torres:2016aa}. In contrast, the phenomena has been well demonstrated using photonic topological insulators using special kinds of metacrystals\cite{Khanikaev:2013aa,Wang:2009aa}.} In this work we show the same phenomenon can be obtained in the proposed tunable re-entrant post structure. To demonstrate this, a source of external signal is incorporated into the centre of the 2D square lattice with an interface between $L$ and $R$ chiral halves. The interface does not have any particular termination at the lattice ends and is equivalent to all other lattice columns and rows of the system. In fact, one of the posts on the media interface serves as an entry port for the signal. This construction can be implemented by putting {a straight antenna microwave port under the corresponding post. A straight antenna port may be thought of as a piece of microwave cable with a straight inner conductor sticking out of the outer ground conductor. This type of port couples to a component of the electrical field parallel to the conductor, which in this case is $E_z$.} Additionally the model is equipped with a simple linear loss mechanism that is identical for each post. The corresponding term is odd time-symmetric that is analogous to the Lorentz force term of a charged particle moving in the fixed external magnetic field {\color{black} that is responsible for unidirectional properties of magnetic systems}.

The energy distribution of the system with external pump and dissipation is shown in Fig.~\ref{topTS}. The figure shows two different order edge modes pumped near the corresponding resonance frequencies. If the system is excited near one of the bulk modes, the wave propagates inside the bulk. The results for edge mode pumping demonstrate propagation of the edge waves of different order along the $L/R$ media interface only in one direction without significant backscattering from {the lattice outer surface. The result is achieved for the system pumped at the resonance frequency of the damped edge mode, i.e. the frequency depending on the resonance frequency of the undamped edge mode and the mode bandwidth, at this frequency the system exhibit the backscattering immune properties\cite{Ningyuan:2015aa, Swinteck:2015aa, Chen:2011aa}.

The phenomenon of one way photon photon transport is important for many practical applications, in particular, as a way to reduce back scattering. Usually the goal is achieved by employing magnetically ordered systems like ferrites\cite{Dietz:2007aa,Anderson2016,Ningyuan:2015aa}, although the time reversal symmetry breaking phenomenon can be observed in paramagnetic spin ensembles as well\cite{Goryachev:2014ab,Goryachev:2014ac}, superconducting qubit based circuit Quantum Electrodynamics\cite{Koch:2010aa}, Hall effect systems\cite{Viola:2014aa}. Thus, the re-entrant post structure provides an important way to achieve this goal in a reconfigurable setup without use of ferromagnetic materials and external magnetic fields. This fact becomes even more important at low temperatures and relatively low frequency where ferromagnetic materials typically  exhibit very high losses preventing them from such applications.

\section{Discussion}

Previous sections demonstrate the possibility of introducing handedness into metamaterials composed of identical non-chiral objects set on a regular grid. The ability to change configuration of the system in any spot gives an advantage of changing the system topology 'on the fly'. Similar systems have been proposed for other applications\cite{Goryachev:2015aa,Goryachev:2015ab} and adds no extra complexity. High localisation of edge modes together with its reconfigurability may be exploited to address certain areas of of the system coupling to some foreign object, i.e. magnetic crystals or qubits thus implementing multi purpose random access systems, e.g. memories. The other possible applications inspired by the reconfigurability property is topological signal processing \cite{topproc} and sensing\cite{Bi:2006aa}. 
 
 The flexibility to introduce chiral properties come at the cost of post redundancy: every minimal block of the lattice, a post molecule, consists of four posts increasing dimensionality of the problem. Moreover, to design more than four bulk bands, one needs to increase the number of posts within a molecule. Although, in principle it is possible to make this number variable by connecting some posts of a molecule to the opposite wall and removing this post from the interaction within the molecule. Thus, the proposed approach is able to extend the idea of Programmable Cavity Arrays\cite{Goryachev:2015ab} to include topological features and make possible related microwave domain signal processing.

\section*{Acknowledgement}

This work was supported by the Australian Research Council grant number CE110001013.


%

\end{document}